\documentclass[fleqn,10pt]{wlscirep}
\title{Evolutionary dynamics of N-person Hawk-Dove games}

\author[1]{Wei Chen}
\author[2]{Carlos Gracia-L\'azaro}
\author[3,1]{Zhiwu Li}
\author[4]{Long Wang}
\author[2,5,6]{Yamir Moreno}
\affil[1]{School of Electro-Mechanical Engineering, Xidian University, Xi'an 710071, China}
\affil[2]{Institute for Biocomputation and Physics of Complex Systems (BIFI), Universidad de Zaragoza, 50018 Zaragoza, Spain}
\affil[3]{Institute of Systems Engineering, Macau University of Science and Technology, Taipa, Macau}
\affil[4]{Center for Systems and Control, College of Engineering, Peking University, Beijing 100871, China}
\affil[5]{Department of Theoretical Physics, Faculty of Sciences, Universidad de Zaragoza, Zaragoza 50009, Spain}
\affil[6]{Institute for Scientific Interchange (ISI), Torino, Italy}


\begin{abstract}
In the animal world, the competition between individuals belonging to different species for a resource often requires the cooperation of several individuals in groups. This paper proposes a generalization of the Hawk-Dove Game for an arbitrary number of agents: the N-person Hawk-Dove Game. In this model, doves exemplify the cooperative behavior without intraspecies conflict, while hawks represent the aggressive behavior. In the absence of hawks, doves share the resource equally and avoid conflict, but having hawks around lead to doves escaping without fighting. Conversely, hawks fight for the resource at the cost of getting injured. Nevertheless, if doves are present in sufficient number to expel the hawks, they can aggregate to protect the resource, and thus avoid being plundered by hawks. We derive and numerically solve an exact equation for the evolution of the system in both finite and infinite well-mixed populations, finding the conditions for stable coexistence between both species. Furthermore, by varying the different parameters, we found a scenario of bifurcations that leads the system from dominating hawks and coexistence to bi-stability, multiple interior equilibria and dominating doves. 
\end{abstract}
\begin{document}

\flushbottom
\maketitle

\thispagestyle{empty}

\section*{Introduction}

Evolutionary Game Theory (EGT) provides a theoretical framework to model Darwinian competition that
has been widely used to study evolving populations of lifeforms in
biology and, particularly, for the study of cooperative behavior in
animals\cite{smith1973lhe,smith1982evolution,hofbauer1998evolutionary,vincent2005evolutionary,axelrod2006evolution}. In this
context, most of the research has focused on two-person games such as the Prisoner’s Dilemma\cite{axelrod1987evolution},
the Stag Hunt\cite{pacheco2009evolutionary} and the Hawk-Dove
Game (HDG)\cite{hauert2004spatial,doebeli2005models}, 
which describe conflicting situations where some individuals 
profit from selfishness to the detriment of others. From these games, the Prisoner's dilemma reproduces
the worst possible scenario for cooperation, in which an agent always gets the highest individual benefit
by not cooperating. The Stag Hunt is a coordination game that describes
a conflict between safety and cooperation: while the greatest benefit for both agents is obtained when
both cooperate, cooperation implies a risk, because cooperating against a non cooperative agent provides
the lowest benefit. The HDG is an anti-coordination game with two possible strategies:
cooperation or defection. While the best strategy against a cooperator is defection, the highest cost is paid
by defecting against a defector. 

The above games or dilemmas have greatly contributed, despite their simplifications, to a better understanding of what could be plausible mechanisms for the emergence and sustainability of cooperative behavior. However, 
many real-life situations involve collective decisions made by a group, rather than by only two
individuals. In this sense, Public Goods Game (PGG)
capture situations in which some individuals (free-riders) can benefit from other ones (cooperators)
without paying any cost, that is, without suffering the decrease of fitness associated with the
production of the common good\cite{perc2013evolutionary}. In the classical PGG, the contributions of cooperators to a common pool
are enhanced and equally shared among cooperators and defectors, regardless their contribution\cite{sigmund2010calculus}. 

In addition to the PGG, there is a large set of possible multi-agent
situations that can be addressed through other games. In this regard, the Snowdrift Game (SG) is a model
of conflict in which there is a task to be done. The basis underlying the conflict is that defectors
benefit from cooperators without paying a cost for doing the task, but without cooperators the task would not be
accomplished\cite{sugden1986economics}. When only two individuals are involved, the situation is described by the standard
SG. In this case, both agents are faced with a common task. There are three possible outcomes: i) No one cooperates, and
hence the task is not achieved with the consequent damage to each one. ii) Both individuals cooperate,
and both benefit, sharing the work associated with the task. iii) If only one individual cooperates,
both benefit from the task done despite the fact that the cooperator has paid the entire cost associated to the task\cite{smith1982evolution}.
If the benefit is greater than the cost, the ordering of payoffs coincides with that of the HDG. This means that
in the case of only two individuals involved, both HDG and SG are equivalent. When more than two agents are
involved, the conflict underlying the Snowdrift problem can be addressed through an N-person generalization of the 
standard SG\cite{zheng2007cooperative}. The N-person SG considers a sample consisting of N agents and a task to be done. Every agent, regardless of whether he contributes to the
task or not, will receive a fixed benefit if the task is performed by one or more agents within the sample. The total cost
of performing the task is equally shared among those who perform it (the cooperators), while the defectors
benefit without paying any cost\cite{chan2008evolution,souza2009evolution,ji2011effects,santos2012dynamics,sui2015evolutionary}. On the other hand, the situation in the Hawk-Dove problem is of a very different nature compared to that underlying the Snowdrift. In the Hawk-Dove problem there is not a task to be done, but a resource susceptible to be shared. While
in the Snowdrift problem every individual benefits from the task, if accomplished, in the Hawk-Dove
problem only a kind of strategists (cooperators or defectors) benefits from the resource,
excluding the opponent strategists from the distribution \cite{hauert2004spatial,doebeli2005models,tomassini2006hawks, voelkl2010hawk}. This different nature materializes
when there are more than two individuals involved; in this general case, the mathematical formulation differs drastically
between both situations, which call for a new approach to deal with the Hawk-Dove problem for many
individuals.

Here, we propose a generalization of HDG for N agents, hereafter N-person HDG, with the aim of analyzing
how the dynamics is affected when considering group interactions. In the N-person HDG, two kinds of strategists (hawks and doves)
compete for a resource $R$. While hawks 
are willing to fight to get the whole resource, doves wish to share the resource
equally, refusing the resource rather than fighting for it. Accordingly, when a set of individuals face a resource, the hawks fight
among them, paying a cost $c_H$, while doves retreat. Only in the case that there are no hawks competing for the resource, the
doves will share it.  The fitness of the individuals is determined by their payoff when getting involved in an
N-person HDG.  We study analytically the evolution of cooperative behavior
within the framework of the replicator dynamics\cite{hofbauer1998evolutionary,cressman2003evolutionary}, by considering
either a very large or an infinite population of hawks and doves. More concretely, we assume a population of size $Z$, from which groups of size $N$ are randomly sampled; let $n_H$ and $n_D$ denote the numbers of hawks and doves in the sample
respectively, $N=n_H+n_D$. For any given group size $N$, we derive an exact equation for the 
evolution of the fraction of doves in a well-mixed population, which can be solved numerically as a function of
the rest of the parameters. We first study the infinite size limit $Z\rightarrow \infty$, and show the
existence of a critical value $c_{cv}$ for $c_H$ above which cooperation is sustainable.
While in the region $c_H>c_{cv}$, hawks and doves coexist in the stationary state, for $c_H<c_{cv}$
the dynamics leads to the extinction of the doves, resulting in a population of only hawk strategists.

Furthermore, as a refinement of the model, we consider the situation in which the doves aggregate to
defend the resource, not at all cost, but only in the case they are present in sufficient number to expel
the hawks, i.e., there is a minimum threshold $T$ that assures them to succeed. In that
case, after expelling the hawks, paying a cost $c_D$ for it, doves share the resource
without fighting among themselves. Otherwise, i.e., if the number of
doves is below the threshold, they retreat and the hawks fight among themselves. It is worth noticing that although both hawks and doves can fight, a hawk will fight in any situation (unless all his opponents flee), while
doves only fight against opposite strategists, and provided they are allowed to expel them. This refinement of the
model, hereafter N-person HDG with threshold (HDG-T), captures the stress between gregarious and asocial behaviors. We
show that, depending on the values of $T,c_D,c_H$, the dynamics drives the system towards either one of the absorbing
mono-strategic states (all hawks or all doves), or towards an interior equilibrium in which both strategies coexist. Subsequently,
we study the dynamics in finite populations, finding results compatible with those obtained
for the infinite size limit $Z\rightarrow \infty$.

\section*{Results}

\subsection*{Preliminary concepts}

As mentioned earlier, the standard HDG is a two-player-two-strategies game that captures a conflict
of interests in which the most advantageous strategy
is the opposite to that of the opponent. There are two possible strategies, namely, hawkish behavior ($H$), 
that represents fighting, and dove behavior ($D$), that represents sharing. When faced with a common
resource $R$, two doves equally share the resource obtaining each one $R/2$,
while two hawks fight for it, paying a cost $c$ that represents the damage caused by the fight,
and therefore obtaining each one $(R-c)/2$. If a hawk meets a dove, the hawk obtains the whole resource
$R$, while the dove retreats and gets nothing. The game can be expressed by means of its
payoff matrix, where rows represent focal player's strategies, columns represent
opponent's strategies, and the corresponding matrix element is the payoff received
by the focal player:

\begin{table}[ht]
\centering
\begin{tabular}{|c|c|c|}
\hline
strategy & H & D \\
\hline
H & $(R-c)/2$ & $R$ \\
\hline
D & $0$ & $R/2$ \\
\hline
\end{tabular}
\end{table}

\subsection*{N-person HDG}

The model we present here constitutes a generalization of the two-person HDG to an N-person game. Assume
a sample of size $N\geq2$ with two kinds of strategists, the aggressive ones $-$ like the hawks$-$, and the
cooperators $-$ the doves. Let $R$ be a resource such as food or water. As pointed out before, when the population is made up by all doves, they share the resource equally avoiding conflict. If doves meet hawks, all doves escape without fighting, while the remaining hawks fight for the resource paying a given cost. Their respective payoffs are:

Hawks' Payoff: 
\begin{equation}
P_H=\frac{R-(n_h-1)c_H}{n_h} \qquad  n_H>0
\label{HpayoffHDG}
\end{equation}

Doves' payoff: 
\begin{equation}
P_D=
\begin{cases}
\frac{R}{N}& n_H=0\\
0& n_H>0
\end{cases}
\label{DpayoffHDG}
\end{equation}
where $n_H$ and $n_D$ represent the numbers of hawks and doves in the sample, respectively, and $c_H$
the cost of injury. Thus, $(n_H-1)c_H$ represents the total injury suffered
by all the hawks. It should be noted that doves
only get payoffs when there are no hawks in the sample. For $N=2$, the model recovers
the standard two-person HDG. In order to study the effect of cost and sample
size and composition, $R$ is conventionally assumed to be 1.

Consider a very large population $Z\rightarrow \infty$, composed of a fraction $x$ of doves,
the rest ($1-x$) being hawks. Social interactions take place in sample groups
of size $N$, which are randomly selected from the whole population. The average
fitness of hawks $f_H(x)$ and doves $f_D(x)$ is determined according to a binomial
sampling, as shown in the methods section. We study the population dynamics in terms of
the replicator equation\cite{hofbauer1998evolutionary}, according to which the
time evolution of $x$ is given by:
\begin{equation}
\dot{x}=x(f_D(x)-\langle f(x)\rangle) \; \; \; , 
\label{replicatorEquationResults}
\end{equation}
where $\dot{x}$ represents the gradient of selection and $\langle f(x)\rangle$ stands for the average fitness of the whole population.

\begin{figure}[ht]
\begin{center}
\includegraphics[width=\linewidth]{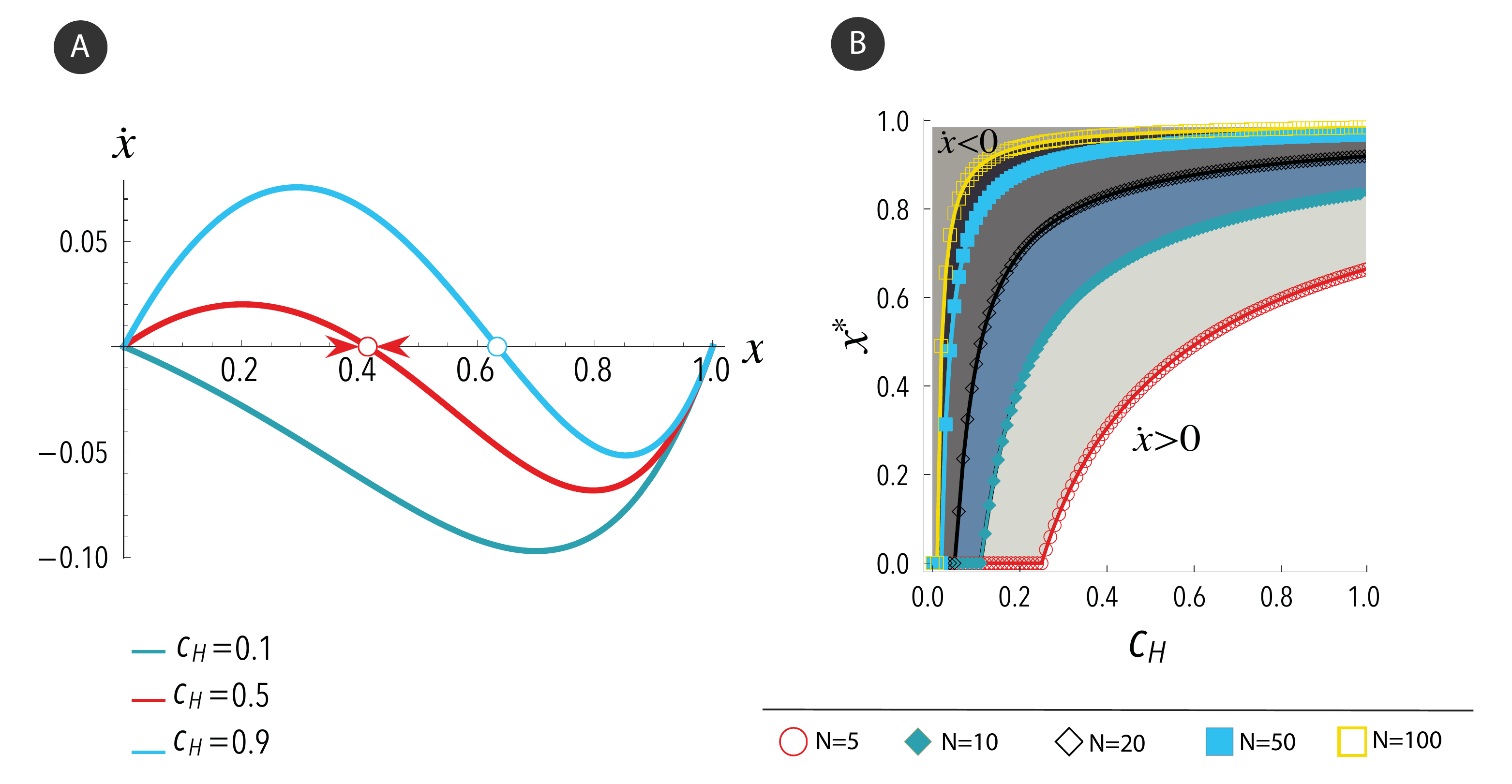}
\caption{\textbf{Gradient of selection and equilibria of the N-person Hawk-Dove Game in infinite populations.} (a) Gradient of selection $\dot{x}$ in the N-person HDG as a function
of the frequency $x$ of doves, for a sample size of $N=5$. Different colors stand for different costs
$c_H=0.1, 0.5, 0.9$. Circles represent the interior attractors
indicating the coexistence of hawks and doves in the steady state. (b) 
Equilibria of the N-person Hawk-Dove Game in infinite populations
for different values of cost $c_H$. $x^*$ represents the equilibrium fraction of doves in the population. There is a critical
value $c_{cv}=1/(N-1)$ for the cost $c_H$ beyond which hawks and doves coexist. 
The resource is taken as unity $R=1$. See the main text for further details.
} 

\label{fig1}
\end{center}
\end{figure}

Figure \ref{fig1}(a) shows the gradient of selection $\dot{x}$ as a function
of the density $x$ of doves, for different values of $c_H$ and a simple size of $N=5$. As shown, for
a small cost ($c_H=0.1$), the gradient of selection is always negative and
doves fail to survive regardless of any initial condition. Nevertheless, 
as the internal conflict among the hawks grows, and with it its associated cost
$c_H$, there appears a critical value, beyond which an internal equilibrium point
indicates the coexistence of doves and hawks in the steady state. Furthermore, setting $f_D(x)=f_H(x)$
and $x=0$, the minimum value of $c_H$ for stable equilibrium to survive is $c_{cv}=R/(N-1)$, being
$c_{cv}$ the critical value. Figure \ref{fig1}(b) shows the
fraction of doves $x^*$ in the equilibrium as a function of the cost $c_H$, for different
sample sizes $N$. As shown, the equilibrium fraction of doves increases with increasing
cost of hawks and increasing sample size. The latter follows from the fact that the greater the 
sample size, the greater the number of hawks involved in the fight. In general, a higher competition
of hawks (either due to an increase in sample size or to an increase in hawks' cost) entails a decrease in their average fitness, and thus the equilibrium frequency of
doves increases. As $c_H>c_{cv}$ implies $x^*>0$, 
it follows that $c_{cv}$ determines the value for the transition from a mono-species state to coexistence.

At this point, let us assume a finite population of size $Z$, composed by $k$ doves
and ($Z-k$) hawks. As in the case of infinite populations, interactions take place
in sample groups of size $N$, but in this case the mean fitness of hawks and doves is
determined in accordance with a multivariate hypergeometric sampling, see the section Methods. Regarding the dynamics, we assume a stochastic birth-death process driven by a pairwise
Fermi-like rule\cite{roca2009evolutionary}. According to this rule,
at each time step, two individuals ($u,v$) are randomly selected. The
probability that a descendant of $v$ replaces $u$ is given by an increasing function of the payoff difference:
\begin{equation}
p=\frac{1}{1+exp[-w(f_u-f_v)]} \;\;\; ,
\label{FermiLikeRule}
\end{equation}
where 
the parameter $w$ controls the intensity of selection. Low $w$ represents high noise
and, correspondingly, weak selection
pressure.

\begin{figure}[ht]
\begin{center}
\includegraphics[width=\linewidth]{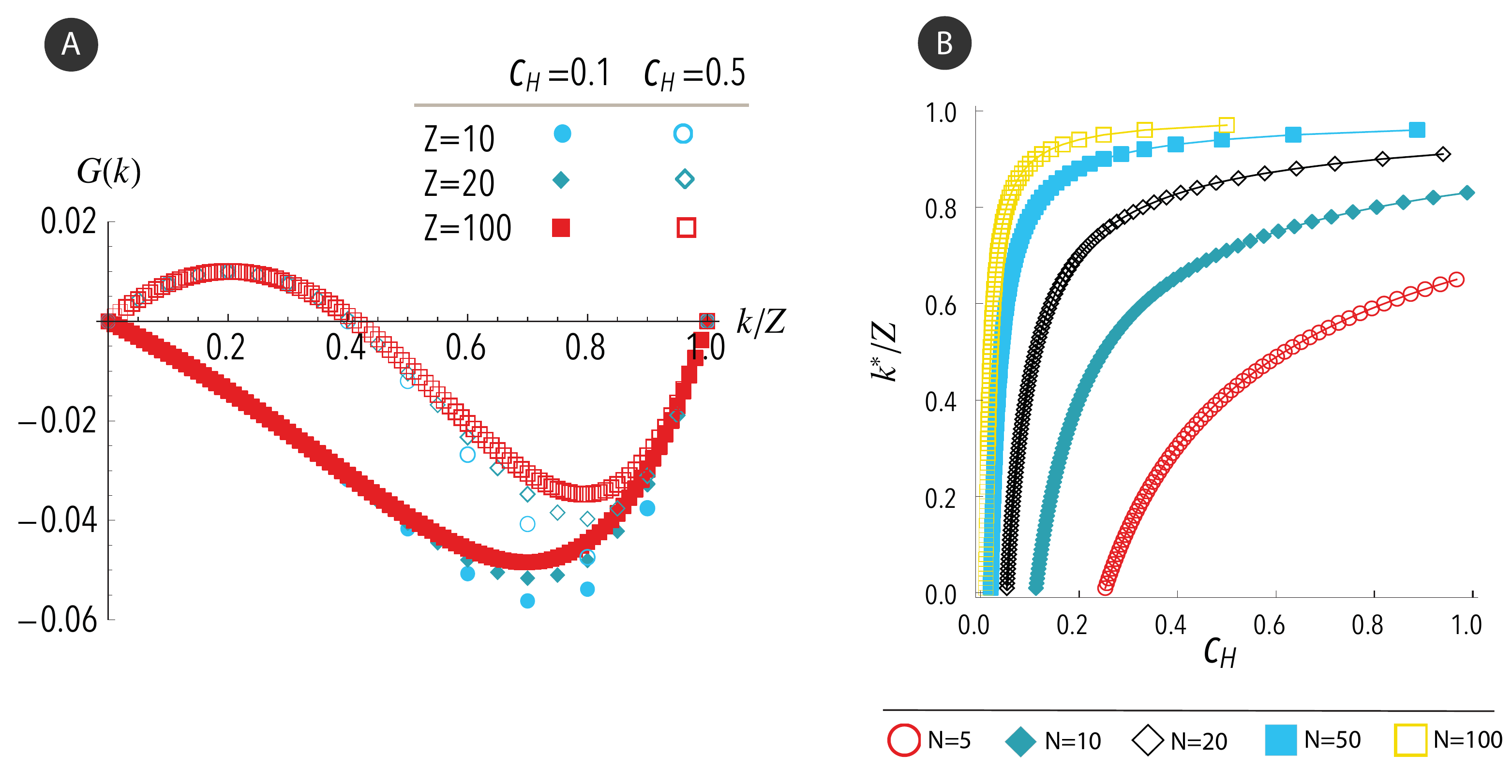}
\caption{\textbf{Gradient of selection and equilibria of the N-person Hawk-Dove Game in finite populations.} (a) Gradient of selection $G(k)$ in the N-person HDG as a function of the discrete fraction $k/Z$ of doves, for different population sizes $Z$ and
a sample size of $N=5$. Different shapes of symbols represent different population sizes $Z=10, 20, 100$,
while solid (\textit{resp.}, empty) symbols represent a cost
of  $c_H=0.1$ (\textit{resp.}, 0.5). (b) Equilibria of the N-person Hawk-Dove Game in finite populations
for different values of cost $c_H$, and a population
size of $Z=100$. 
Symbols denote the required values of $c_H$ for which there exists an equilibrium at $k/Z$.
Note that, for finite populations, $k/Z$ takes discrete values.
Different symbols represent different sample sizes $N$. Other parameters
are $R=1$, $w=1$.} 
\label{fig2}
\end{center}
\end{figure}

Figure \ref{fig2}(a) shows the gradient of selection $G(k)$ describing the
evolutionary dynamics of N-person HDG in finite populations as a function of the fraction
of doves $k/Z$, for different population sizes $Z$ and costs $c_H$ and a sample size of $N=5$. It should
be noted that, for finite populations, the fraction of doves $k/Z$ is not a continuous variable, but a
discrete one. It is shown that, as in the previously discussed case of infinite populations, 
for a small cost $c_H$ the gradient of selection is always negative $G(k)<0$,
and therefore doves become extinct, regardless of their initial fraction in the population. For higher cost
values $c_H$, there exists a unique internal root $k^*$ making $G(k)=0$. Therefore, a stable equilibrium arises with
increasing $c_H$ and doves are promoted. In addition, both the gradient of selection $G(k)$ and the
internal stable equilibrium $k^*$ increase slightly as the group size $Z$ increases, which implies that large populations facilitate the advantage of doves. Fig. \ref{fig2}(b) depicts the equilibrium fractions
of doves as a function of the cost $c_H$, for different sample
sizes $N$ and a fixed population size $Z=100$. Note that these results are compatible with those obtained for finite populations. Thus, when increasing $c_H$, the regime in which the hawks dominate changes to a regime in which both species coexist. Moreover, increasing the sample size enhances the fitness of doves and, consequently, less hawks' cost $c_H$ is required for the existence of doves.

\subsection*{N-person HDG-T}

In the above described N-person HDG, doves are assumed to be non-aggressive. As a
refinement of the model, now we assume that doves can aggregate to protect the resource
and thus avoid being plundered by hawks. This situation is very common in nature, where
animals such as buffaloes aggregate to fight against predators. In this revised model, 
N-person HDG-T, we assume a threshold $T$. If the
fraction of doves exceeds or equals the threshold ($n_D/N\geq T$), the doves expel the hawks, paying
a cost $c_D$ for it, and then share the resource equally without fighting among themselves. Otherwise, 
if the number of doves is below the threshold ($N_D/N<T$), doves retreat and hawks 
fight among themselves for the resource. The respective payoffs of doves and hawks in the 
N-person HDG-T are:

Hawks' payoff: 
\begin{equation}
P_H=
\begin{cases}
0& n_D/N\geq T\\
\frac{R-(n_H-1)c_H}{n_H} & n_D/N< T
\end{cases}
\end{equation}

Doves' payoff:
\begin{equation}
P_D=
\begin{cases}
\frac{R-n_Hc_D}{n_D}& n_D/N\geq T\\
0& n_D/N< T
\end{cases}
\end{equation}
where $n_H$ (\textit{resp.}, $n_D$) stands for the number of hawks (\textit{resp.}, doves) in the
sample of size $N$, $c_D$ for the cost of doves
to protect the resource from being plundering by hawks, and $c_H$ for the cost of injury within
hawks. Notice, once again, the different nature of both costs: while
$c_H$ refers to intra-specific competition, $c_D$ refers to an interspecific (intergroup)
conflict cost. For $T=1$ the previous N-person HDG model is recovered.

\begin{figure}[ht]
\begin{center}\includegraphics[width=\linewidth]{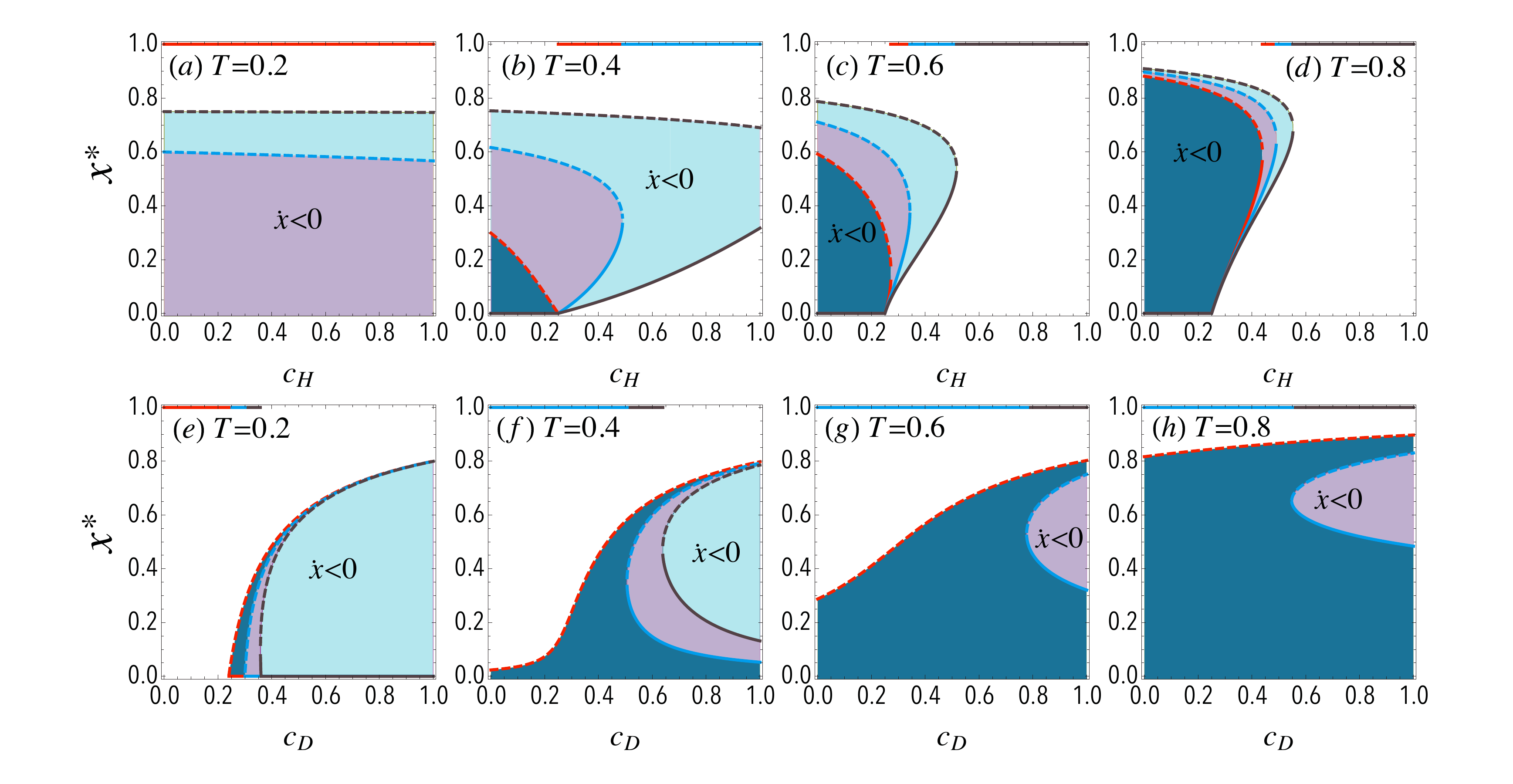}
\caption{\textbf{Equilibria of the N-person Hawk-Dove Game with threshold in infinite populations.} Fraction of doves  $x^*$ at the equilibrium for a
sample size of $N=5$.  Solid (\textit{resp.}, dashed) line corresponds to stable (\textit{resp.}, unstable) equilibrium.  Top panels 
represent $x^*$ as a function of the hawks' cost $c_H$. Different color lines stand for different
doves' costs: red, blue and gray lines represent $c_D=0.2,0.5,0.8$ respectively.
The bottom panels represent $x^*$ as a function of the doves' cost $c_D$. Different color lines stand for different
hawks' costs: red, blue and gray lines represent $c_H=0.2,0.5,0.8$ respectively.
Shaded areas denote a negative gradient of
selection. (a,e), (b,f), (c,g) and (d,h) panels correspond to thresholds $T=0.2,0.4,0.6,0.8$ respectively.
The resource is taken as unity $R=1$. See the main text for further details.}
\label{fig3}
\end{center}
\end{figure}

As in the previous model, for the case of a very large population ($Z\rightarrow \infty$)
the population dynamics is studied according to the replicator equation. Figure
\ref{fig3} represents the frequency of doves in the equilibrium $x^*$ for different thresholds and costs
in the limit of an infinite population. Upper
panels show how the intra-specific conflict cost of hawks $c_H$ impacts the evolution of doves
for different thresholds $T=0.2,0.4,0.6,0.8$ and for different fixed values of doves' cost $c_D$. As shown in panel (a), when the threshold is
small ($T=0.2$), indicating that doves are more resourceful in intergroup competition, a slight cost of doves ($c_D=0.2$) drives the population to evolve into a full-dove state and no interior equilibrium exists. When $c_D$ is increased, an unstable equilibrium emerges and divides the system into two basins of attraction regardless of $c_H$. Both full-dove state and full-hawk state can be reached from different initial conditions: for an initial frequency of doves $x_0>x^*$, the
hawks vanish ($f_D>f_H$), but for $x_0<x^*$, the doves are the ones that vanish ($f_D<f_H$), indicating that both
absorbing states are stable equilibria and the proposed game is transformed into a coordination game. Panels
(b-d) show the behavior for greater thresholds $T>0.2$. In these cases, for a small $c_H$, an interior unstable equilibrium appears, whose value decreases for larger $c_H$, indicating that increasing $c_H$ enlarges the basin of attraction of the full-dove steady state. With further increasing $c_H$, multiple interior equilibria can occur and a new stable equilibrium emerges. Finally, for even higher values of $c_H$, both unstable and stable equilibria converge to a saddle-node bifurcation beyond which the dynamics always leads to a full-dove state. 

Bottom panels of Fig. \ref{fig3} show the effects of the intergroup conflict cost of doves $c_D$. When
the threshold is small ($T=0.2$), as shown in panel (e), $x^*=1$ is the only steady state for low values of
$c_D$, 
which means that the dynamics always leads to a full-dove state regardless of the initial condition. With
the increase of $c_D$, an unstable equilibrium appears and the dynamics changes from
dominating doves to bi-stability. However, for $c_H=0.8$, a saddle-node bifurcation emerges as $c_D$ is
increased, but the stable equilibrium only exists for a very small range of values of $c_D$, resulting in
a transcritical bifurcation that leads to bi-stability.  Panels (f-h) show the behavior for greater
thresholds $T>0.2$. In these cases, a small $c_H$ leads to an interior unstable
equilibrium, while a large $c_H$ results in the domination of doves. Yet for a moderate $c_H$, multiple
interior equilibria including an unstable equilibrium and a stable equilibrium drive the system to
either the domination of doves or the coexistence of both species under different initial conditions.

\begin{figure}[ht]
\begin{center}
\includegraphics[width=\linewidth]{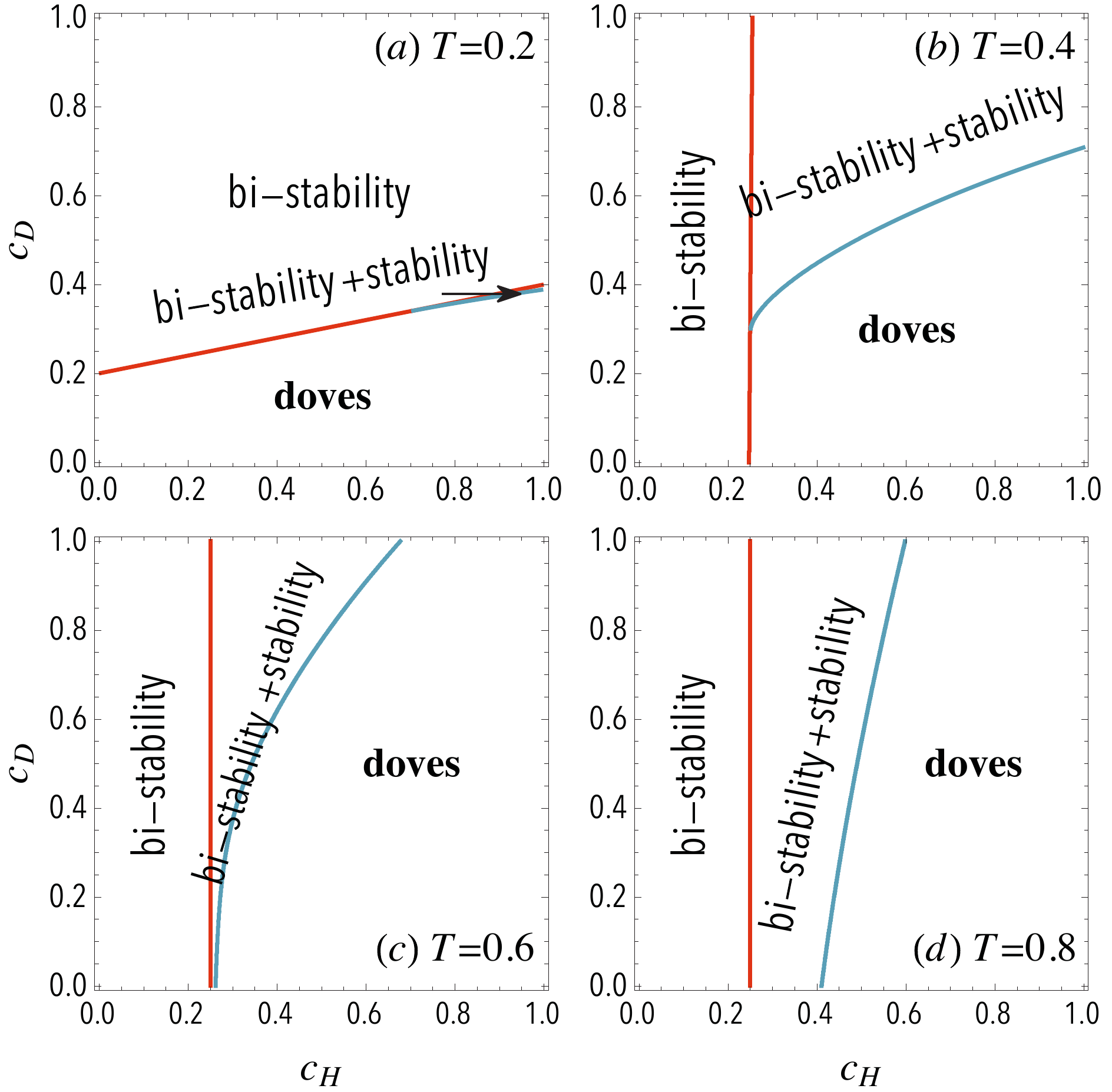}
\caption{\textbf{Phase diagrams for the N-person Hawk-Dove Game with threshold.} 
Diagrams show the different regimes in the N-person HDG-T for a
sample size of $N=5$ and an infinite population. Different panels correspond to different thresholds $T = 0.2, 0.4, 0.6, 0.8$ respectively. Within the {bi-stability} regime both full-dove state and full-hawk state can be reached from
different initial conditions, while the \textit{bi-stability+stability} regime corresponds to
coexistence state together with full-dove state. In the \textit{doves} regime the dynamics always leads to a full-dove state. The resource is taken $R=1$.} \label{fig4}
\end{center}
\end{figure}

Figure \ref{fig4} shows a natural classification of the dynamics of N-person HDG-T that depicts three basic phase regimes. The condition that favors hawks is met when $P_D(N)-P_H(N-1)<0$, indicating that a single hawk in a group of $N-1$ doves is better off than a dove in a full-dove group. For $T=1$ this condition becomes $P_D(N)-P_H(N-1)=R/N-R<0$, which is always met, and hawks do not vanish (shown in Fig. \ref{fig1}(b)). Besides, the condition for a successful invasion of doves is obtained when a single dove in a group of $N-1$ hawks has a higher income than a full hawkish group, namely, $P_D(1)-P_H(0)>0$. For $T=1/N$ this condition becomes $c_H-Nc_D+R>0$, see the red line in Fig. \ref{fig4}(a). Below this line, multiple interior equilibria are found $-$though in a narrow region$-$, as well as a full-dove state. However, for $T\geq 2/N$, one gets that $c_H>R/(N-1)$ from the relation $P_D(1)-P_H(0)>0$, which provides a condition for the emergence of stable equilibria. Results in Figure \ref{fig4} highlight the possible transitions between three different phases as a function of the costs and the threshold $T$. Consequently,
both increasing $c_H$ and decreasing $c_D$ facilitate the fixation of the doves in the population.

Furthermore, $T$ also has critical effects on the evolutionary dynamics. For instance, by increasing it, one can find regions where dominating doves give rise to bi-stability, and other changes such as from multiple interior equilibria
to dominating doves, or from bi-stability to multiple interior equilibria and even to dominating doves. Additionally, increasing $T$ suppresses doves even for a small $c_H$, while there is an optimal $T$ that facilitates doves. For $c_H>R/(N-1)$, increasing $T$ not only strengthens the intergroup competitiveness of doves $-$ there are less hawks to fight with, thus, the cost is smaller$-$ when $n_D/N\geq T$, but also weakens the in-group competitiveness of hawks, since the decrease in the maximum number
of hawks to win over doves results in less in-group injury when $n_H/N>T$. As a consequence, an optimal
$T$ is found to promote doves. Particularly, for $c_H\leq R/(N-1)$, the payoff of a single dove in a group of $N-1$ hawks is lower than the payoff of a hawk in a full hawkish group since $P_D(1)-P_H(0)>0$, and therefore
doves are suppressed. On the other hand, a small $c_H$ enhances the competitiveness of hawks
in intergroup conflict. As a consequence, increasing $T$ reduces the in-group conflict of hawks
and finally favors hawks. Regarding the effects of sample size, regardless of costs, increasing $N$
suppresses doves for $T=1/N$, since the basin of attraction of the hawkish state increases, whereas
doves are favored for $T=1$ since stable equilibrium increases. This behavior
is mainly attributed to the impact of sample size on in-group competition and resource allocation, 
benefiting the incomes of victorious agents from the reduced reduction.

\begin{figure}[ht]
\begin{center}
\includegraphics[width=\linewidth]{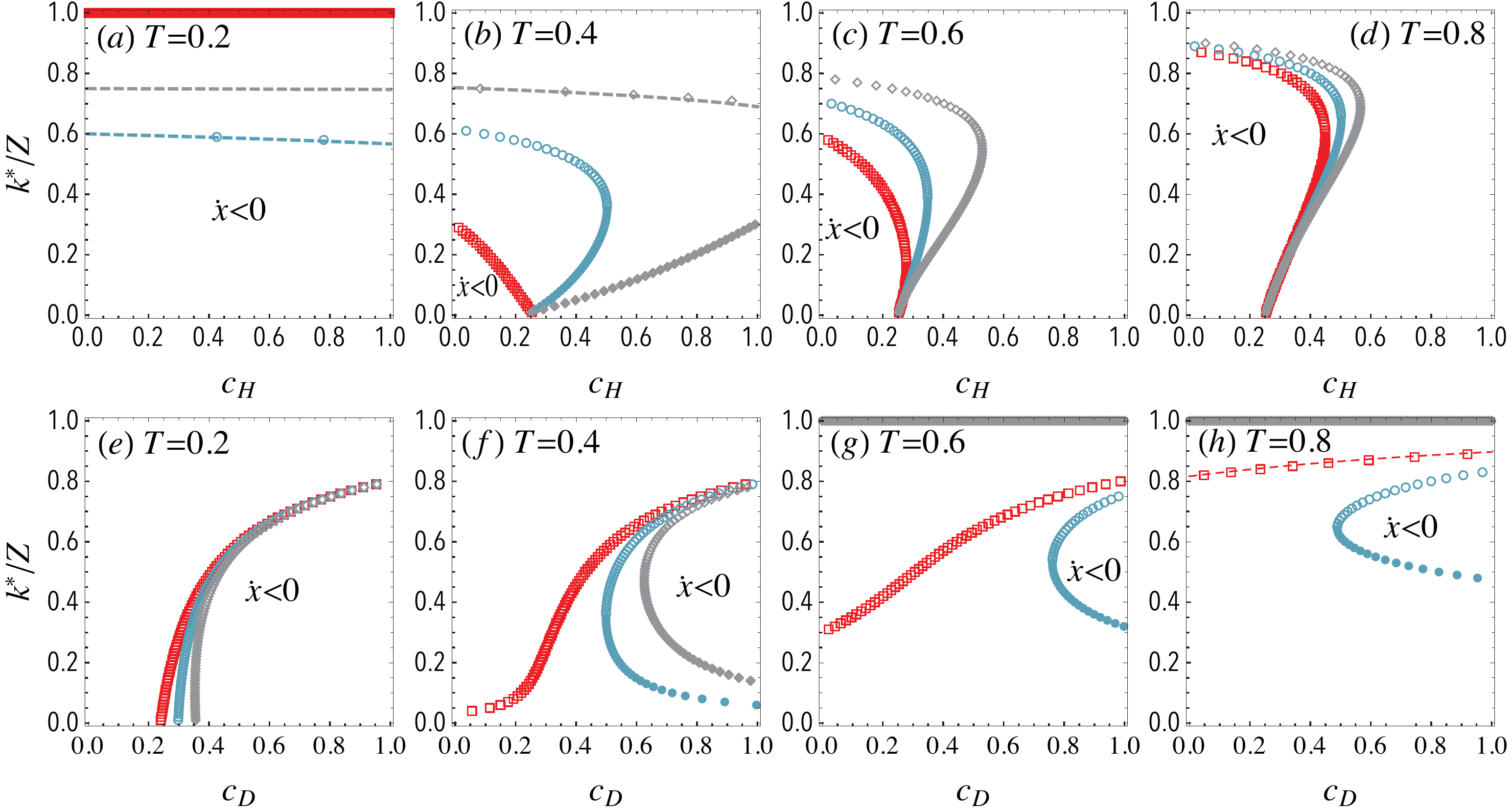}
\caption{\textbf{Equilibria of the N-person Hawk-Dove Game with threshold in finite populations.} Fraction of doves $k^*/Z$ in the equilibrium as a
function of the hawks' cost $c_H$ (top) and doves' cost $c_D$ (bottom), for a sample size of $N=5$ in a finite population of size $Z=100$. Solid (dashed) points represent stable (bistable) points, whereas dashed lines mark the regions beyond which $f_D>f_H$ and below which $f_D<f_H$. In top panels, 
different colored symbols represent different
doves' costs: red, blue and gray lines correspond to $c_D=0.2,0.5,0.8$, respectively.
In the bottom panels, red, blue and gray symbols represent hawks' costs $c_H=0.2,0.5,0.8$, respectively.
Other parameters are $R=1$, $w=1$.} \label{fig5}
\end{center}
\end{figure}

As for finite populations, Fig. \ref{fig5} shows the equilibrium fraction of doves $k^*/Z$ as a function of the costs $c_H$ and $c_D$, for different sample sizes $N$ and a fixed population size $Z=100$. The interior roots of $G(k)$, varying with different costs and thresholds, are presented when the roots exist. These results are very similar to those corresponding to infinite populations
shown in Fig. \ref{fig3}. Particularly, since successful invasions can be directly obtained from $P_D(k)$ and $P_H(k)$, which are independent of binomial or hypergeometric samplings, the phase diagrams are extremely similar to those corresponding to infinite populations shown in Fig. \ref{fig4}. We have also explored the effects of threshold and sample size, which in finite populations are not qualitatively different from those reported before for infinite populations. Moreover, by exploring the stationary
distribution of doves, it is also found that both increasing $c_H$ and decreasing $c_D$ promote
doves. Finally, although the equilibrium points show a small dependence on the population size $Z$, it vanishes as $Z$ grows, finding that for sufficiently high values of $Z$ the dynamics is independent of the population size, which means that for large enough populations ($Z/N\geq 100$) the infinite population limit is a fair approximation that captures the dynamics of the system.

\section*{Discussion}

In summary, this work proposes a general framework of multi-player Hawk-Dove game that generalizes the traditional symmetric two-person Hawk-Dove game. Two different kinds of species, denoted as hawks and doves respectively, represent the competition to survive of hawkish and cooperative behaviors through group interactions. The hawkish behavior, a more competitive trait, wins intergroup conflict, but also bears the cost associated with intra-group conflict, whereas the cooperative behavior, overcoming intra-species conflict, may lose in interspecies conflict unless the fraction of doves exceeds a threshold $T$. By implementing the proposed model in well-mixed infinite and finite populations, we found that increasing the intra-group competition cost of hawks, as well as decreasing the intergroup competition cost of doves, promotes the evolution of doves. Furthermore, the threshold plays a key role in the model. As $T$ decreases from $T=1$, the phase regimes changes from dominating hawks and stability to bi-stability, multiple interior equilibria and dominating doves. Specially, if the intra-group competition cost of hawks is greater than $R/(N-1)$, an optimal threshold is found to optimize doves, otherwise increasing threshold suppresses them.

Although two-person Hawk-Dove Games can mathematically be regarded as a Snowdrift Game, this equivalence breaks down when generalizing it to N persons. An N-person Snowdrift Game characterizes real-world situations in which a task needs to be done by cooperating, with the consequent benefit for the group, whereas the proposed model describes the competition for a resource between two species. The N-person Hawk-Dove here presented thus conceptualizes the dilemma that being an aggressive type can reward from interspecies competition but also incurs a high cost in terms of intra-species conflict, while cooperators aggregate and share the resource. We also hypothesize that, if the present model were to be applied to a real context in nature, the threshold needed for successful preservation of the doves type would be very high. Indeed, a high threshold is more realistic than the assumptions on which either doves obtain nothing unless no hawks are in the group or a low threshold that implies a fiercer intra-species conflict. Furthermore, group interactions can not be regarded as a set of independent pairwise encounters. Finally, we mention that in previous studies of a two-person Hawk-Dove Game, both an heterogeneous topology and different update rules have been proven to have an impact on cooperation \cite{tomassini2006hawks, voelkl2010hawk}. It would thus be of further interest to explore how such factors change our findings for N-person hawk-dove games, and therefore gain more insights into our current understanding of cooperative behavior in social dilemmas.

\section*{Methods}
\subsection*{Evolutionary dynamics in infinite populations}

Consider a very large well-mixed population $Z\rightarrow \infty$, composed of a fraction $x$ of doves
and a fraction ($1-x$) of hawks. Sample groups of size $N$ are randomly selected from
the population. Following a binomial sampling \cite{hauert2006synergy}, the mean fitness of hawks in the population is given by:
\begin{equation}
f_H(x)=\sum_{i=0}^{N-1} \tbinom{N-1}{i}x^i(1-x)^{N-1-i}P_H(i) \; \; \; ,
\label{fitnessH}
\end{equation}
whereas the average fitness of doves is:
\begin{equation}
f_D(x)=\sum_{i=0}^{N-1} \tbinom{N-1}{i}x^i(1-x)^{N-1-i}P_D(i+1) \; \; \; .
\label{fitnessD}
\end{equation}

The time evolution of $x$ is given by the replicator equation\cite{hofbauer1998evolutionary}:
\begin{equation}
\dot{x}=x(f_D(x)-\langle f(x)\rangle)=x(f_D(x)-(xf_D(x)+(1-x)f_H(x)))=x(1-x)(f_D(x)-f_H(x)) \; \; \; ,
\label{replicatorEquation}
\end{equation}
where $\langle f(x)\rangle$ represents the average fitness of the whole population. From this equation, it follows that the equilibria satisfy:
\begin{equation}
f_D(x^*)-f_H(x^*)=0 \;\;\;.
\end{equation}

As an illustrative example, let us consider the case $T=N$, corresponding to the N-person HDG. According to equations (\ref{HpayoffHDG},\ref{fitnessH}), the mean fitness of hawks is given by:
\begin{equation}
f_H(x)=\frac{(R+c_H)(1-x^N)}{N(1-x)}-c_H \;\;\;,
\end{equation}
while from equations (\ref{DpayoffHDG},\ref{fitnessD}) it follows that the mean fitness of doves is:
\begin{equation}
f_D(x)=\frac{Rx^{N-1}}{N}  \;\;\;.
\end{equation}

Hence, we get:
\begin{equation}
f_D(x)-f_H(x)=\frac{c_Hx^N+Rx^{N-1}-Nc_Hx+Nc_H-R-c_H}{N(1-x)} \;\;\;,
\end{equation}
and the replicator equation (\ref{replicatorEquation}) can be written as:
\begin{equation}
\dot{x}=\frac{x}{N}(c_Hx^N+Rx^{N-1}-Nc_Hx+Nc_H-R-c_H)
\end{equation}

\subsection*{Evolutionary dynamics in finite populations}

In this subsection, we address the evolutionary dynamics of N-person HDG in finite populations of
size $Z$. Unlike the case of infinite populations, where sampling followed a binomial distribution,
the groups in finite populations are given by a multivariate hypergeometric sampling\cite{young1993evolution,kandori1993learning}. Let $k$ be the number
of doves in the total population. Accordingly, the number of hawks will be $Z-k$. Now,
the fraction of doves ($k/Z$) is not a
continuous variable, but a discrete one. The average fitness of hawks
is thus given by:
\begin{equation}
f_H(k)=\sum_{i=0}^{N-1} \frac{\tbinom{k}{i}\tbinom{Z-k-1}{N-i-1}}{\tbinom{Z-1}{N-1}}P_H(i) \;\;\; ,
\end{equation}
whereas the average fitness of doves is:
\begin{equation}
f_D(k)=\sum_{i=0}^{N-1}\frac{\tbinom{k-1}{i}\tbinom{Z-k}{N-i-1}}{\tbinom{Z-1}{N-1}}P_D(i+1) \;\;\; .
\end{equation}

Assuming the Fermi-like rule\cite{roca2009evolutionary}, at each elementary time step two individuals are
chosen at random from the population. If they are of different species, then the probabilities that a dove replaces a hawk, and the opposite scenario, are given, respectively, as 

\begin{equation}
p_{D\rightarrow H}=\frac{1}{1+exp[-w(f_D(k)-f_H(k))]} \;\;\; ,
\end{equation}
and
\begin{equation}
p_{H\rightarrow D}=\frac{1}{1+exp[-w(f_H(k)-f_D(k))]} \;\;\; .
\end{equation}

Consequently, the probabilities that the finite population increases ($T^+$) or decreases ($T^-$) by one dove are\cite{traulsen2006stochastic}:
\begin{equation}
T^\pm(k)=\frac{k}{Z}\frac{Z-k}{Z}\frac{1}{1+e^{\mp w(f_D(k)-f_H(k))}} \;\;\; ,
\end{equation}
and therefore, the gradient of selection in finite populations is given by:
\begin{equation}
G(k)=T^{+}(k)-T^{-}(k)=\frac{k}{Z}\frac{Z-k}{Z}tanh(\frac{w}{2}(f_D(k)-f_H(k)))
\end{equation}

\bibliography{nplayersHD}

\section*{Acknowledgements (not compulsory)}

This work was supported in part by the National Natural Science Foundation of China under Grant No. 61374068 and
the Science Technology Development Fund, MSAR, under Grant No. 066/2013/A2. C. G. L. and Y. M. acknowledge support from the Government of Arag\'on, Spain through a grant to the group FENOL, by MINECO and FEDER funds (grant FIS2014-55867-P) and by the European Commission FET-Open Project Ibsen (grant 662725). 

\section*{Author contributions statement}

W. C., C. G. L., Z. L., L. W. and Y. M. conceived the study, C. W. performed the simulations, W. C., C. G. L., Z. L., L. W. and Y. M. analyzed the results and W. C., C. G. L., Z. L., L. W. and Y. M. wrote the manuscript. All authors approved the final version.

\section*{Additional information}


The corresponding author is responsible for submitting a \href{http://www.nature.com/srep/policies/index.html#competing}{competing financial interests statement} on behalf of all authors of the paper.

\end{document}